\documentclass[twocolumn,aps,pra,superscriptaddress,showpacs]{revtex4}
\usepackage{graphicx}
\usepackage{amsmath}
\usepackage{amssymb}

\begin{document}

\title{Highly efficient generation of single-mode photon pairs using a crystalline whispering gallery mode resonator}

\author{Michael F\"{o}rtsch}
\affiliation{Max Planck Institute for the Science of Light, G\"{u}nther-Scharowsky-Str. 1, Bau 24, 91058, Erlangen, Germany}
\affiliation{Institut f\"{u}r Optik, Information und Photonik, University of Erlangen-Nuremberg, Staudtstr. 7/B2, 91058, Erlangen, Germany}
\affiliation{SAOT, School in Advanced Optical Technologies, Paul-Gordan-Str. 6, 91052 Erlangen}

\author{Gerhard Schunk}
\affiliation{Max Planck Institute for the Science of Light, G\"{u}nther-Scharowsky-Str. 1, Bau 24, 91058, Erlangen, Germany}
\affiliation{Institut f\"{u}r Optik, Information und Photonik, University of Erlangen-Nuremberg, Staudtstr. 7/B2, 91058, Erlangen, Germany}
\affiliation{SAOT, School in Advanced Optical Technologies, Paul-Gordan-Str. 6, 91052 Erlangen}

\author{Josef U. F\"{u}rst}
\affiliation{Max Planck Institute for the Science of Light, G\"{u}nther-Scharowsky-Str. 1, Bau 24, 91058, Erlangen, Germany}
\affiliation{Institut f\"{u}r Optik, Information und Photonik, University of Erlangen-Nuremberg, Staudtstr. 7/B2, 91058, Erlangen, Germany}

\author{Dmitry Strekalov}
\affiliation{Max Planck Institute for the Science of Light, G\"{u}nther-Scharowsky-Str. 1, Bau 24, 91058, Erlangen, Germany}

\author{Thomas Gerrits}
\affiliation{National Institute of Standards and Technology, 325 Broadway, Boulder, CO 80305, USA}

\author{Martin J. Stevens}
\affiliation{National Institute of Standards and Technology, 325 Broadway, Boulder, CO 80305, USA}

\author{Florian Sedlmeir}
\affiliation{Max Planck Institute for the Science of Light, G\"{u}nther-Scharowsky-Str. 1, Bau 24, 91058, Erlangen, Germany}
\affiliation{Institut f\"{u}r Optik, Information und Photonik, University of Erlangen-Nuremberg, Staudtstr. 7/B2, 91058, Erlangen, Germany}
\affiliation{SAOT, School in Advanced Optical Technologies, Paul-Gordan-Str. 6, 91052 Erlangen}

\author{Harald G. L. Schwefel}
\affiliation{Max Planck Institute for the Science of Light, G\"{u}nther-Scharowsky-Str. 1, Bau 24, 91058, Erlangen, Germany}
\affiliation{Institut f\"{u}r Optik, Information und Photonik, University of Erlangen-Nuremberg, Staudtstr. 7/B2, 91058, Erlangen, Germany}

\author{Sae Woo Nam}
\affiliation{National Institute of Standards and Technology, 325 Broadway, Boulder, CO 80305, USA}

\author{Gerd Leuchs}
\affiliation{Max Planck Institute for the Science of Light, G\"{u}nther-Scharowsky-Str. 1, Bau 24, 91058, Erlangen, Germany}
\affiliation{Institut f\"{u}r Optik, Information und Photonik, University of Erlangen-Nuremberg, Staudtstr. 7/B2, 91058, Erlangen, Germany}

\author{Christoph Marquardt}
\affiliation{Max Planck Institute for the Science of Light, G\"{u}nther-Scharowsky-Str. 1, Bau 24, 91058, Erlangen, Germany}
\affiliation{Institut f\"{u}r Optik, Information und Photonik, University of Erlangen-Nuremberg, Staudtstr. 7/B2, 91058, Erlangen, Germany}

\begin{abstract}
We report a highly efficient source of narrow-band photon pairs based on parametric down-conversion in a crystalline whispering gallery mode resonator. Remarkably, each photon of a pair is strictly emitted into a single spatial and temporal mode, as witnessed by Glauber’s autocorrelation function. We explore the phase-matching conditions in spherical geometries, and  determine the requirements of the single-mode operation. Understanding these conditions has allowed us to experimentally demonstrate a single-mode pair-detection rate of $0.97 \cdot 10^6$ pairs/s per mW pump power per 20 MHz bandwidth without the need of additional filter cavities.
\end{abstract}


\maketitle
Pure single photons and quantum-correlated photon pairs are indispensable in fundamental tests of quantum mechanics \cite{Aspect:1982,Weihs:1998,Christensen:2013,Giustina:2013} as well as in quantum information technologies such as, quantum computing \cite{Knill:2001}, quantum enhanced measurements \cite{Giovannetti:2004}, and quantum communication \cite{Gisin:2007}. The visionary concepts of these fields have driven much scientific progress in developing photon sources based on various physical implementations such as semiconductor quantum dots \cite{Michler:2000,Santori:2002}, trapped atoms \cite{Kuzmich:2003,Thompson:2006,Wilk:2007}, and four-wave-mixing \cite{Soeller:2010,Srivathsan:2013}. Although these sources have demonstrated excellent properties, they typically emit at a fixed wavelength, which limits their compatibility with other components based on a different physical implementation. 

To be more versatile, one can use spontaneous parametric down-conversion (SPDC) of light for generating photon pairs. In general though, the SPDC process naturally produces broadband, multimode photons \cite{Martinelli:2003}. Lately, there has been much effort and progress into developing SPDC sources that produce \textit{single-mode} broadband quantum states for ultrafast pulses of light \cite{Mosley:2008,Harder:2013}. 
This effort is motivated by the ability to preserve high purity when interfering in quantum optical networks or interconnects.

Efficient interfacing of photon sources with atomic transitions requires not only wavelength tunability, but also  narrow-band emission. This can be achieved in resonator assisted spontaneous parametric down-conversion (RA-SPDC). This process offers the possibility to efficiently generate narrow-band heralded single photons \cite{Ou:1999,Scholz:2009,Fekete:2013,Wolfgramm:2011,Scholz:2009_2}, which are directly compatible with atomic transitions  of quantum memories \cite{Lvovsky:2009}. One remaining challenge with RA-SPDC based systems is efficient photon generation in exactly one spatio-temporal mode, which is often accompanied by additional losses originating from inefficient filtering. Lately, a waveguide based RA-SPDC source was reported, which offers the potential to intrinsically generate single-mode photon pairs \cite{Luo:2013}. 

In this Letter, we demonstrate an alternative approach for a narrow-band photon pair source based on a crystalline whispering gallery mode resonator (WGMR) \cite{Ilchenko:2004,Fuerst:2010_1,Beckmann:2011,Furst:2011} that emits photons in exactly one optical mode. Crystalline WGMR systems offer wide wavelength and bandwidth tuning while the efficiency can be orders of magnitudes higher than in non-monolithic resonators \cite{Fortsch:2013}. The key for the single-mode operation is the physics of the angular phase-matching solutions for the second order parametric process in spherical geometries. Its precise understanding enables us to directly design the phase-matching and allows for a single-mode emission without the need of additional lossy filter cavities. 
We demonstrate an unprecedented pair-detection rate of $0.97 \cdot 10^6$ pairs/s per mW pump power per 20\,MHz bandwidth. To the best of our knowledge, this is the first time that true single-mode operation was witnessed by Glauber's autocorrelation function \cite{Glauber:1963} in combination with a RA-SPDC system. 

Phase-matching in spherical geometries is based on the conservation of both the energy and, in contrast to plane waves, the orbital angular momentum of the pump (p), signal (s), and idler (i) photons \cite{Kozyreff:2008,Fuerst:2010_2}.The efficiency of this process depends on the interaction energy $\sigma$ that is expressed by the overlap integral:
\begin{eqnarray}
	\label{eq:interaction_energy}
	\sigma = \int \chi^{(2)} \Psi_s(r,\theta,\phi) \Psi_i(r,\theta,\phi) \Psi^*_p(r,\theta,\phi) dV.
\end{eqnarray}
Here, $\Psi_{p,s,i}$ represent the time-independent wavefunctions of the interacting photons, which are eigenfunctions of the surrounding geometry, and $\chi^{(2)}$ is the effective quadratic nonlinearity. 
\begin{figure}[htbp]
\includegraphics[width=8cm]{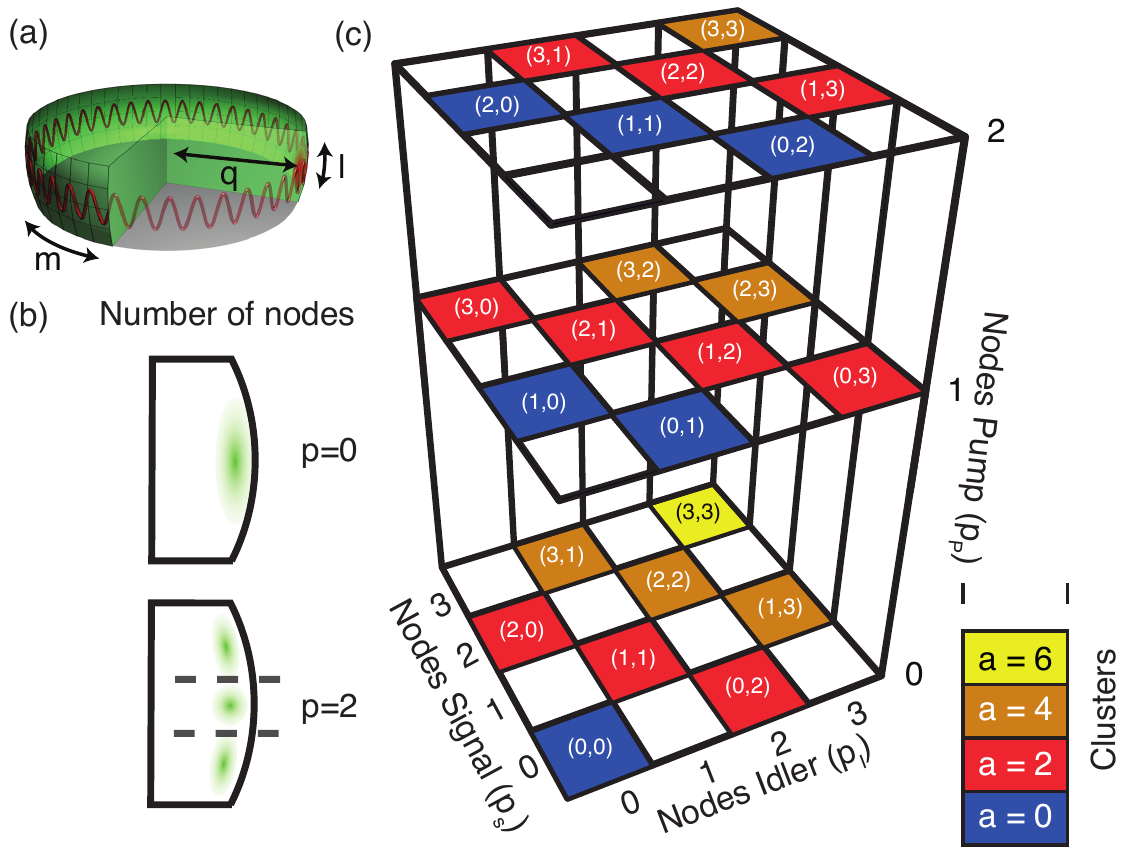}
\caption{\label{fig:poss_phase_matching_angular_001}\textbf{Possible angular phase-matching solutions}: (a) An illustration of a WGMR explaining the mode indices $q,m,l$. (b) A cross-section of the intensity distribution inside a WGMR for two different angular mode numbers $p$. (c) An illustration of the angular phase-matching solutions. Each layer in this diagram represents a subset of the possible solutions for a given pump mode $p_p$. The individual cluster numbers $a$ are encoded in color.}
\end{figure}
If the spherical objects is large compared to the optical wavelength, the integration can be performed separately for the radial and an angular part. The radial part is described by spherical Bessel functions $j_l(k_q r)$ and the angular part is described by spherical harmonics $Y_{lm}(\theta,\varphi)$. 

Here, $m$, $l$, and $q$ are the azimuthal, polar and, radial mode number, which characterize each eigenmode of the WGMR completely (see Fig.~\ref{fig:poss_phase_matching_angular_001}(a)). Introducing the angular mode parameter $p = l - |m|,\;p\geq0$ is experimentally convenient \cite{Gorodetsky:1994} as $p$ represents the number of nodes of a WGM along the polar direction (see Fig.~\ref{fig:poss_phase_matching_angular_001}(b)). 

One can express the angular part of the non vanishing overlap integral Eq.~\ref{eq:interaction_energy} using the indices of the spherical harmonics with the following selection rules:
\begin{subequations}\label{eq:selction_rules}
	\begin{equation}\label{eq:selection_m}
	m_p = m_s + m_i,
	\end{equation}
	\begin{equation}	\label{eq:selection_l_1}
	\left | l_s - l_i \right | \leq l_p \leq l_s + l_i,
	\end{equation}
	\begin{equation}\label{eq:selection_l_2}
	l_p + l_s + l_i = {2\mathbb{Z}}.
	\end{equation}
\end{subequations}
In contrast to plane waves, the WGMR eigenfunctions are not multiplicative. This means a product of two WGMR eigenfunctions is not orthogonal to a third one except in respect to the azimuthal mode number $m$. This leads to a large variety of possible phase-matching solutions. 
Here, we investigate how the phase-matching solutions depend on the angular mode number $p$.

Expressing Eq.~\ref{eq:selection_m} in terms of $p$ leads to: 
\begin{eqnarray}
	\label{eq:phase_matching_002}
	\underbrace{(l_s + l_i - l_p)}_{=2\mathbb{N}_{0}} - \underbrace{(p_s + p_i - p_p)}_{\equiv a}  = 0.
\end{eqnarray}
According to Eq.~\ref{eq:selection_l_2}, $(l_s + l_i - l_p)$ is an even positive integer, thus $(p_s + p_i - p_p)\equiv a$ is an even positive integer as well. 
For analyzing the possible angular phase-matching solutions, we will fix the orbital number $l_x$ and consider only  the $p$-dependent part. Because $p\geq0$, there exist a well-defined set of angular mode number combinations leading to the same number $a\in2\mathbb{N}_0$, which we will call \textit{cluster number}.
A subset of these possible angular solutions is illustrated in Fig.~\ref{fig:poss_phase_matching_angular_001}c.  In the following we will show that clusters represent nearly frequency-degenerate phase-matching solutions that can make it difficult to achieve single-mode operation of the source.

In Fig.~\ref{fig:poss_phase_matching_angular_001}(c) each horizontal layer represents the possible phase-matching solutions for a pump mode with a fixed angular mode number $p_p$. This representation reflects the typical experimental situation where the pump mode, and thereby the layer of the diagram, is selected. The individual clusters are highlighted using different colors. Depending on the selected pump mode each cluster contains a different number of possible phase-matching solutions. The number of modes $k$ comprising one cluster is determined by the angular pump mode number $p_p$ and the cluster number $a$ as $k = p_p + a +1$. Therefore for an equatorial pump mode $(p_p=0)$ the first cluster $a=0$ contains exactly one possible phase-matching solution. This feature offers the possibility to realize a pair photon source that emits photons in one single optical mode. 

\begin{figure*}[htbp]
\includegraphics[width=12cm]{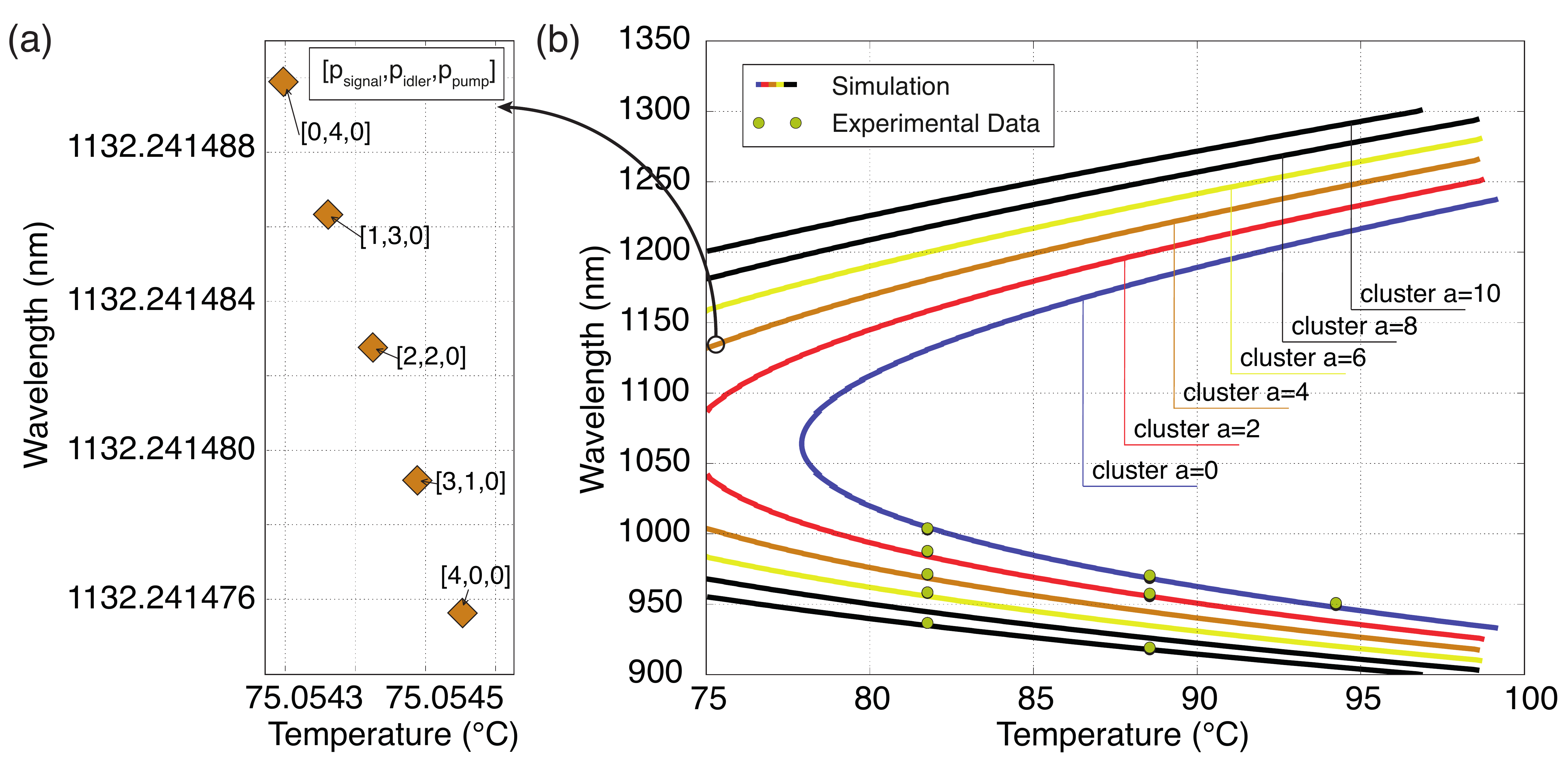}
\caption{\label{fig:spectral_distribution}\textbf{Spectral distribution}: Numerical simulation of the spectral distribution for a subset of possible clusters. (a) A close-up of the cluster $a=4$ highlighting the dense spectral distribution of the possible phase-matching solutions within one cluster. (b) The spectral distribution of the first six possible of clusters phase-matched with a fundamental pump mode ($q_p=1, p_p=0$). The yellow dots represent the measured data points. }
\end{figure*}

In order to estimate the spectral distribution of the individual clusters and the frequency difference between possible phase-matching solutions within one cluster, we numerically simulate the energy conservation $\omega_s + \omega_i = \omega_p  $. The individual frequencies are calculated using the dispersion relation for spheroidal geometries \cite{Gorodetsky:2006}. Following the above phase-matching analysis, true single-mode operation requires and equatorialpump mode. Therefore we restrict our numerical simulations to $p_p=0$, corresponding to the bottom layer in Fig.~\ref{fig:poss_phase_matching_angular_001}(c). 
The possible mode combinations within one cluster $a$ differ only by a few megahertz (Fig.~\ref{fig:spectral_distribution}a). The individual clusters however are well separated by several nanometers from each other in the spectrum (Fig.~\ref{fig:spectral_distribution}b).

This spectral separation, in combination with the fact that the first cluster ($a=0$) comprises exactly one phase-matching solution for an equatorial pump mode ($p_p=0$), allows for an easy and efficient detection of narrow-band single photons in exactly one optical mode. 

\begin{figure}[htbp]
	\centering\includegraphics[width=8cm]{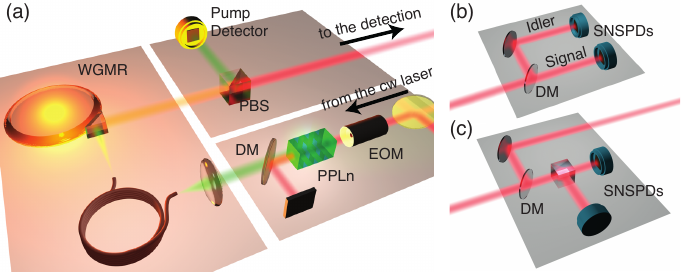}
	\caption{\label{fig:experiment}\textbf{Experimental setup} (a) Temporal pulses are prepared by sending the initial continuous wave Nd:YAG laser through an electro-optical modulator (EOM). The 1064\,nm pulses are frequency-doubled using a periodically poled lithium niobate (PPLN) crystal and sent to an optical fiber. Narrow-band photon pairs are generated via SPDC in a crystalline whispering gallery mode resonator (WGMR). The residual pump light is removed with a polarizing beam-splitter (PBS). (b) For investigating the cross-correlation and the pair-detection function, the signal and the idler photons are separated using a dichroic mirror (DM) and individually recorded using two super conducting nanowire single photon detectors (SNSPDs). (c) For investigating the autocorrelation function the two output ports of a beamsplitter in the signal arm are recorded with two SNSPDs.}
\end{figure}

To investigate our theoretical predictions we use an experimental setup (Fig.~\ref{fig:experiment}) similar to \cite{Fortsch:2013}. 
The pump source is a continuous wave (cw) Nd:YAG laser, emitting light at 1064\,nm. By using a electro-optical modulator (EOM), we carve pulses with a temporal length of 100\,ns from the initial cw light wtih a repetition rate of 4\,MHz. The light pulses are subsequently frequency doubled to 532\,nm using a periodically poled lithium niobate (PPLn) crystal and are coupled to an optical fiber. This cascaded pump pulse generation prevents an unwanted cw background in the 532\,nm pump light due to imperfections of the EOM.
A GRIN lens at the end of the fiber focuses the pump light onto the backside of a diamond prism, which is placed in close vicinity to the resonator. 
The distance between the prism and the resonator can be varied using a piezo that allows for controlling the optical tunneling between the beam's footprint at the prism and the resonator. 
The heart of our setup is a WGMR manufactured from 5.8\% Mg-doped z-cut lithium niobate crystal (LiNbO$_{3}$) with a rim radius of $R\approx1.61$~mm and a small polar radius of $r\approx0.4$~mm. The quality factor of the resonator at the pump wavelength of 532\,nm was measured to be Q$\approx 3\cdot10^{7}$.
We control the phase-matching between the extraordinary polarized pump and the ordinary polarized down-converted signal and idler photons (natural Type I phase-matching \cite{Fuerst:2010_1}) by stabilizing the temperature of the WGMR. 
By choosing an optical pump power of 750\,nW we operated the system in the linear gain regime \cite{Fortsch:2013} far below the optical threshold  \cite{Fuerst:2010_2}.
Signal and idler, as well as residual pump light, exit the resonator via the diamond prism and are subsequently separated by using a polarizing beam splitter (PBS). 
The residual pump light is monitored using a PIN detector for locking the frequency of the pump laser to the WGMR. The frequency of the non-degenerate signal and idler photons are further separated using a dichroic mirror and are directed to the individual detection setups. 
The optional bandpass filter which was used to optically isolate the first cluster ($a=0$) was constructed using a 1000\,nm longpass filter and a 1075\,nm short pass filter. The short pass filter was necessary because the dichroic mirror had a non-negligible reflectivity of $2\%$ for the idler photons. 
The detectors are fiber-coupled, WSi-based superconducting nanowire single-photon detectors (SNSPDs) operated at 150\,mK in a commercial dilution refrigerator. Each detector has a timing jitter of around 120\,ps (FWHM) and a system detection efficiency $\eta\geq$80\% at a wavelength of 1550\,nm \cite{Marsili:2013}.

\begin{figure}[htbp]
\centering\includegraphics[width=8cm]{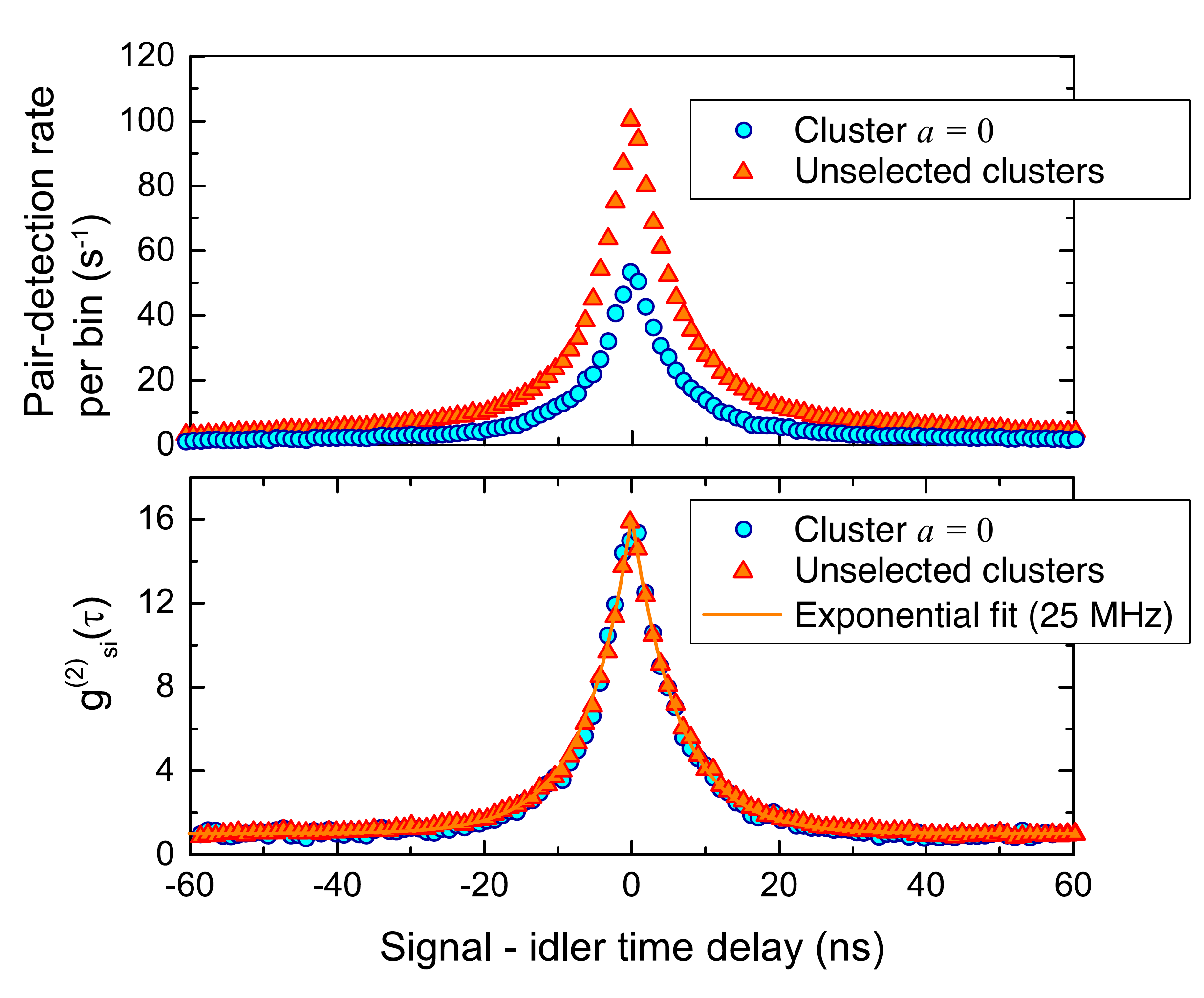}
\caption{\label{fig:pair_and_cross}\textbf{Cross-correlation function}: (a) Pair-detection rate with (blue circles) and without (orange triangles) the optical bandpass filter. (b) Normalized cross-correlation function measured with (blue circles) and without (orange triangles) the bandpass-filter. The bin width in both cases was set to 1\,ns. The bandwidth and the peak value are determined from an exponential fit (solid line) to the experimental data.}
\end{figure}

In order to confirm the spectral distribution of the different clusters predicted by the numerical simulation (Fig.~\ref{fig:spectral_distribution}b), we first investigate the signal wavelength using an optical spectrum analyzer (OSA). To meet the sensitivity of the OSA the experiment was operated above the optical pump threshold \cite{Fuerst:2010_2}. The measurement was performed at three different resonator temperatures while ensuring that the radial number $q$ and the angular number $p$ of the pump mode were kept unchanged. In good agreement with our theoretical prediction, we find that the individual clusters are spectrally separated by at least 10\,nm and are therefore easy to separate with an optical band-pass filter.  

To demonstrate the true single-mode character of the first cluster ($a=0$), two measurements were performed. First, we investigated the stream of photons from the resonator without selecting any cluster.
Second, the first cluster ($a=0$) was optically isolated by inserting the optional bandpass filter in the signal arm. 
In both measurements the quantum state of the detected stream of photons was analyzed by evaluating the pair-detection histogram and Glauber's correlation function \cite{Glauber:1963}:

\begin{eqnarray}
	\label{eq:correlation_func_quantum}
	g^{(2)}_{xy} (\tau) = \frac{\langle \hat{E}_{x}^{(-)}(t) \hat{E}_{y}^{(-)}(t+\tau) \hat{E}_{y}^{(+)}(t+\tau) \hat{E}_{x}^{(+)}(t)\rangle}{\langle \hat{E}_{x}^{(-)}(t) \hat{E}_{x}^{(+)}(t) \rangle \langle \hat{E}_{y}^{(-)}(t) \hat{E}_{y}^{(+)}(t)  \rangle}.
\end{eqnarray}

This measurement investigates the temporal correlations between two optical fields, each represented by a positive field operators $\hat{E}^{\pm}_{xy}(t), (x,y\in{(s,i)})$. Dependent on the selection of the correlated light fields the function is called cross-correlation ($x\neq y$) or autocorrelation ($x=y$) function. 

\textbf{Pair-detection rate} 
For investigating the pair-detection rate $R_c$, the signal and the idler photons are separated using a dichroic mirror. Each beam is then coupled into fiber and sent to an SNSPD (Fig.~\ref{fig:experiment}b). Correlation electronics record a histogram of the time delays separating detection of a signal photon from detection of an idler photon. By normalizing the correlated coincidence counts from the two SNSPDs to the measurement time we receive the pair-detection histograms for the unselected case and the selected first cluster (Fig.~\ref{fig:pair_and_cross}a). 
For a direct comparison between these sets of data, the measurement of the first cluster was corrected for the additional transmission losses of 20\% caused by the band-pass filter. No further losses have been considered for the estimation of the pair-detection rate. 
For the case without selecting a cluster, we find a pair-detection rate of $1.93 \cdot 10^6$ pairs/s per mW pump power per 20\,MHz bandwidth while the pair-detection rate for isolated first cluster ($a=0$) is half as large.

\textbf{Cross-correlation function}
The same experimental configuration (Fig.~\ref{fig:experiment}b) was used to evaluate the cross-correlation function $g^{(2)}_{si}(\tau)$. 
Here, the pair-detection rate $R_c$ between the two SNSPDs is normalized to the product of the signal and idler detector count rates ($R_{s},R_{i}$) and the histogram binning width is set to $\tau_{\text{bin}}$=1\,ns ($g^{(2)}_{si}(\tau) = R_c/(R_s R_i \tau_{\text{bin}})$). 
In general the function is sharply peaked at $\tau=0$  and has a width $\Delta t$ representing the correlation time of the investigated photons. 

In a resonator-enhanced SPDC process, the correlation time is equal to the coherence time of the emitted photons and determined by the resonator's bandwidth $\nu$. Moreover, the maximum of the cross-correlation function is sensitive to the mean photon number $\left\langle n \right\rangle$, and for the low-gain regime is expected to follow $g^{(2)}_{si}(0) = 2 + \frac{1}{\left\langle n \right\rangle}$ \cite{Christ:2011}. This enables us to probe the generated mean photon number that in the low-gain regime, is independent of the underlying mode structure.

As expected, the normalized cross-correlation function for the measurements with and without the bandpass filter are nearly identical for a pump power of 750\,nW. We compare both measurements by using exponential fits of the form $A + B\cdot \exp({- 2 \pi \nu |t|})$ to the experimental data. For both cases we find a mean photon number of $\langle n \rangle = 0.07$ for each detection arm and an optical bandwidth $\nu$ of 26.8 $\pm 0.4$\,MHz. The 95\% confidence bands were calculated using the Levenberg Marquardt fitting routine.

\textbf{Autocorrelation function}
Since the individual photon-number distributions of signal and idler are each expected to follow a thermal distribution \cite{Yurke:1987}, one expects to see the characteristic bunching peak when measuring the autocorrelation function. The maximum of the autocorrelation function is highly sensitive to the number of effective modes $k$ \cite{Eberly:2006} and should scale as  $g^{(2)}_{s s}(0) = 1 + \frac{1}{k}$\cite{Christ:2011}. 
\begin{figure}[htbp]
\centering\includegraphics[width=9.5cm]{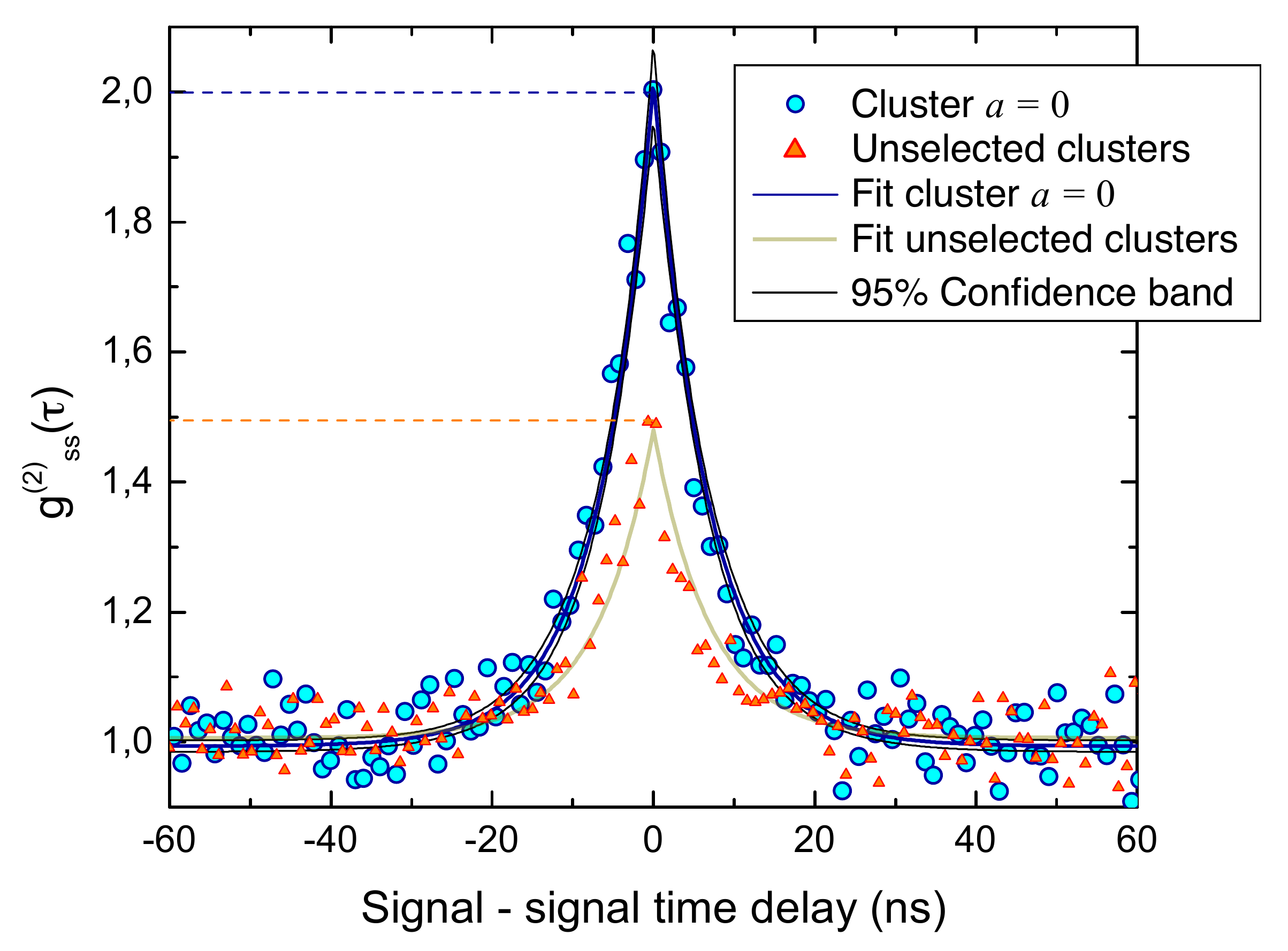}
\caption{\label{fig:Autocorrelation}\textbf{Autocorrelation}: Normalized autocorrelation function with a bin size of 1\,ns. The peak value is determined using an exponential fit (solid curve) to the experimental data (symbols). The 95\% confidence interval is calculated using the Levenberg Marquardt fitting routine.}
\end{figure}
For the autocorrelation function $g^{(2)}_{s s}$ the temporal correlations within one light field, $\hat{E}_{s}(t)$, are measured. To do this, the signal field is split using a 50/50 beamsplitter and correlations between the photons from the two output ports are measured using two SNSPDs (Fig.~\ref{fig:experiment}c). We again compare the measurements with and without the optical bandpass filter using exponential fits to the experimental data. For the unselected case we measure a $g^{(2)}_{s s}(\tau =0) = 1.49$, which corresponds to the detection of two effective modes. When selecting the first cluster with the optical bandpass filter, we measure a $g^{(2)}_{s s}(\tau =0) = 2.01 (1.95 - 2.08)$, which indicates the detection of a pure single-mode thermal state. The 95\% confidence bands were again calculated using the Levenberg Marquardt fitting routine. 

We note that the measured decrease in the pair-detection rate by a factor of two for the selected first cluster coincides well with the factor of two reduction of the effective number of modes. Considering that the experiment was operated using a pulsed pump, this leads us to the conclusion that we have a pair-detection efficiency of $0.97 \cdot 10^6$ pairs/s per mW pump power per 20\,MHz bandwidth for a single spatio-temporal mode.

In summary, we have demonstrated the single-mode selection of a thermal state originating from a highly-efficient pair photon source based on a crystalline whispering gallery mode resonator. In combination with the pulsed operation of the experiment, the detected quantum state can be considered truly single-mode. To the best of our knowledge, this is the first time that a single-mode detection from a resonator enhanced SPDC process has been directly witnessed using Glauber's auto-correlation function. The key to success was understanding the possible phase-matching mode combinations and their spectral distribution for spheroidal objects. These unique phase-matching conditions make it unnecessary to use additional filter cavities and result, to the best of our knowledge, in the highest reported single-mode pair-production rate. In combination with the unique wavelength and bandwidth tuning possibilities, our setup is ready to serve as the single photon source in a large variety of proposed quantum-repeater networks of tomorrow.





\begin{acknowledgements}
We gratefully acknowledge the discussions with Ulrich Vogl, and the support from BMBF grant QuORep.
\end{acknowledgements}


\newpage 
.\newpage

\section{Supplementary Information}
\subsection{Detailed fitting parameters}
\begin{table}[h]
\begin{tabular}{p{1.5cm}|p{1.8cm}|p{1.8cm}|p{1.8cm}}
& A & B & C  \\ 
\hline 
Value & 1.00669 & 15.6172 & 1.6869\,$10^8$ \\ 
95\% LCL & 0.97478 & 15.37992 & 1.64806\,$10^8$\\
95\% UCL & 1.0386 & 15.85447 & 1.72592\,$10^8$
\end{tabular}
\caption{\label{tab:Fitting_value_cross}\textbf{Fitting parameters for cross-correlation}: The values for the fitting of the cross-correlation function based on the equation: 
$g^{(2)}_{\text{si}}(\tau) = A + B \exp{(- C |t|)}$. The 95\% lower confidence limit (LCL) and the upper confidence limit (UCL) were calculated using the Levenberg Marquardt fitting routine.}
\end{table}

\begin{table}
\begin{tabular}{p{1.5cm}|p{1.8cm}|p{1.8cm}|p{1.8cm}}
& A & B & C  \\ 
\hline 
Value & 0.9937 & 1.0248 & 1.416\,$10^8$ \\ 
95\% LCL & 0.9848 & 0.9652 & 1.333\,$10^8$\\
95\% UCL & 1.0026 & 1.0844 & 1.596\,$10^8$
\end{tabular}
\caption{\label{tab:Fitting_value_auto}\textbf{Fitting parameters for autocorrelation}: The values for the fitting of the autocorrelation function based on the equation: 
$g^{(2)}_{\text{ss}}(\tau) = A + B \exp{(- C |t|)}$. The 95\% lower confidence limit (LCL) and the upper confidence limit (UCL) were calculated using the Levenberg Marquardt fitting routine.}
\end{table}

\subsection{Transmission Pump mode}
\label{subs:transmission_pump_mode}
For each measurement presented in the main paper we recorded the transmitted pump light using the pump mode detector. 
By fitting a Lorentzian to the measured pump mode spectrum we observe a linewidth of 78.1($\pm$0.9)\,MHz, which corresponds to a Q(uality) factor of $Q\approx7.2\cdot\,10^6$ for the pump wavelength of $\lambda_p=532$\,nm. The best Q-factor of the resonator was measured to $Q\approx3\cdot\,10^7$ for the 532\,nm pump light. 
\begin{figure}[htbp]
\centering\includegraphics[width=8cm]{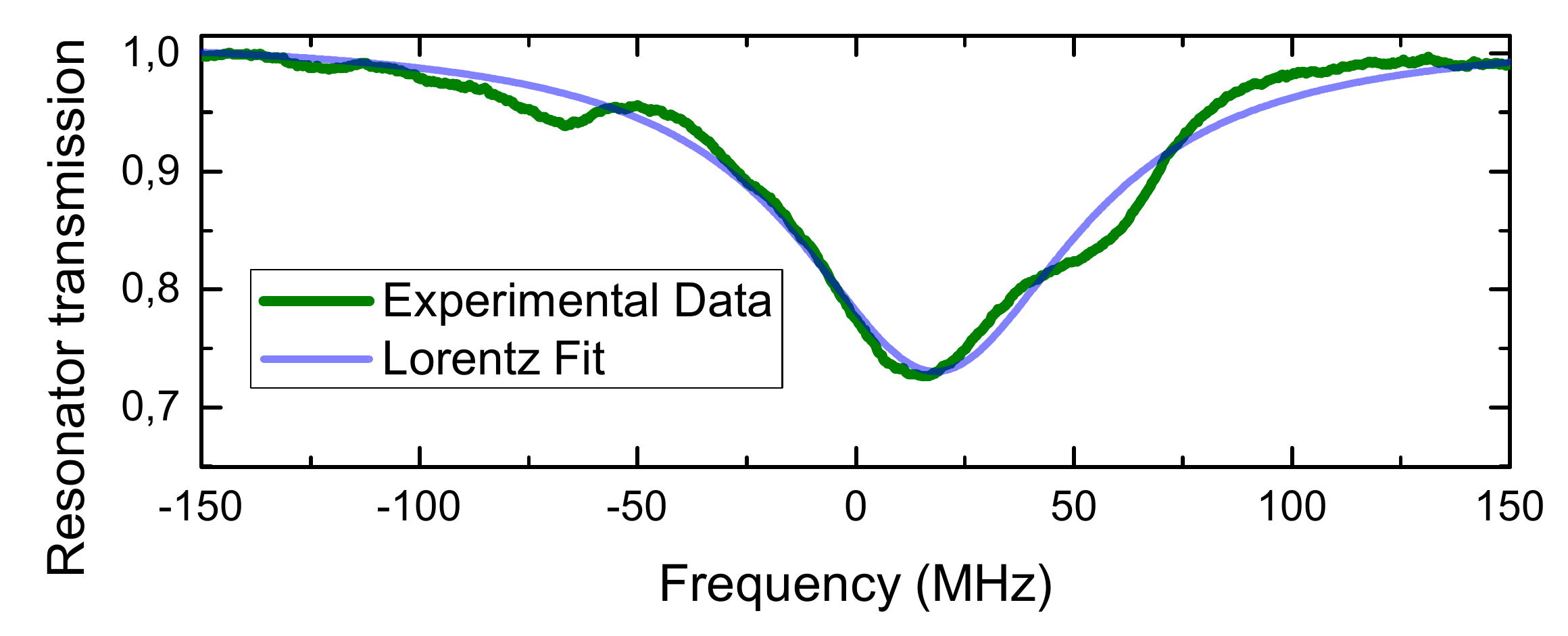}
\caption{\label{fig:transmission_pump_mode}\textbf{Transmission pump mode}: Transmission of the 532\,nm pump mode measured with the pump detector. By fitting a Lorentzian to the measured data we obtain a width of 78.1($\pm$0.9)\,MHz.}
\end{figure}
This result is also supported by the estimated bandwidth of the down-converted light using the cross-correlation function. 
\end{document}